\documentstyle[prd,aps]{revtex}
\input epsf
\newlength{\headroom}
\newlength{\psfigskip}
\renewcommand{\ensuremath}[1]{$#1$}
\renewcommand{\mathrm}[1]{{\rm #1}}
\def\be{\begin{equation}}
\def\ee{\end{equation}}
\def\la{\mathrel{\mathpalette\fun <}}
\def\ga{\mathrel{\mathpalette\fun >}}
\def\fun#1#2{\lower3.6pt\vbox{\baselineskip0pt\lineskip.9pt
  \ialign{$\mathsurround=0pt#1\hfil##\hfil$\crcr#2\crcr\sim\crcr}}}
\def\h3{\ensuremath{^3\mathrm{H}}}
\def\he3{\ensuremath{^3\mathrm{He}}}
\def\4he{\ensuremath{^4\mathrm{He}}}
\def\6li{\ensuremath{^6\mathrm{Li}}}
\def\7li{\ensuremath{^7\mathrm{Li}}}
\def\b7{\ensuremath{^7\mathrm{Be}}}
\def\14n{\ensuremath{^{14}\mathrm{N}}}
\def \yp {\ensuremath{\mathrm{Y}_\mathrm{P}}}
\def \ypm {{\rm Y}_{\rm P}}
\def\e10{\ensuremath{\eta_{10}}}
\newcommand{\omegab}{\Omega_\mathrm{B}}
\newcommand{\deriv}[2]{\frac{\mathrm{d}#1}{\mathrm{d}#2}}
\def\alp{\ensuremath{\alpha}~}
\def\eg{{\it e.g.}}
\def\ie{{\it i.e.}}
\def\etal{{\it et al.~}}
\def\hii{H\thinspace{$\scriptstyle{\rm II}$}~}
\def\popii{Pop\thinspace{$\scriptstyle{\rm II}$}~}
\def\popi{Pop\thinspace{$\scriptstyle{\rm I}$}~}

\def \dtdec {\Delta t_d}
\def \npf {(n/p)_f}  
\begin{document}
\title{Nucleosynthesis in Power-Law Cosmologies}

\author{%
M. Kaplinghat$^{1}$\footnote[3]{Present Address: Dept of Astronomy \& 
Astrophysics, University of Chicago, Chicago, IL 60637}, 
G. Steigman$^{1,2}$ and T. P. Walker$^{1,2}$%
}
\address{%
{\it $^1$Department of Physics, The Ohio State University, Columbus, OH 
43210}\\ 
{\it $^2$Department of Astronomy, The Ohio State University, Columbus, OH 
43210}%
}
\maketitle
%
\begin{abstract}

We have recently considered cosmologies in which the Universal scale
factor varies as a power of the age of the Universe and concluded that
they cannot satisfy the observational constraints on the present age, 
the magnitude-redshift relation for SN Ia, and the primordial element 
(D, \he3, \4he, and \7li) abundances.  This claim has been challenged 
in a proposal that suggested a high baryon density model ($\omegab 
h^2\simeq 0.3$) with an expansion factor varing linearly with time 
could be consistent with the observed abundance of primoridal helium-4, 
while satisfying the age and magnitude-redshift constraints.  In this 
paper we further explore primordial nucleosynthesis in generic power-law 
cosmologies, including the linear case, concluding that models selected 
to satisfy the other observational constraints are incapable of 
accounting for {\it all} the light element abundances.
\end{abstract}

\section{Motivation}
We have studied a class of cosmological models in which the Universal
scale factor grows as a power of the age of the Universe ($a \propto 
t^{\alpha}$) and concluded that such models are not viable since constraints 
on the present age of the Universe and from the magnitude-redshift 
relation favor $\alpha = 1.0 \pm 0.2$, while those from the abundances 
of the light elements produced during primordial nucleosynthesis require 
that $\alpha$ lie in a very narrow range around 0.55~\cite{us}.  Successful 
primordial nucleosynthesis provides a very stringent constraint, requiring 
that a viable model simultaneously account for the observationally inferred 
primordial abundances of deuterium, helium-3, helium-4 and lithium-7.  
For example, if the nucleosynthesis constraint is satisfied, the present 
Universe would be very young; $t_{0} = 7.7~\mathrm{Gyr}$ for a Hubble 
parameter $\mathrm{H}_{0} = 70~\mathrm{kms}^{-1}\mathrm{Mpc}^{-1}$ (or, 
requiring $\mathrm{H}_{0} \leq 54~\mathrm{kms}^{-1}\mathrm{Mpc}^{-1}$ 
for $t_{0} \geq 10~\mathrm{Gyr}$).  

Recently, Sethi \etal \cite{sethi} noted  that cosmologies where the scale 
factor grows linearly with time may produce the correct amount of \4he 
{\it provided that} the Universal baryon fraction is sufficiently large.  
At first this result might seem counter-intuitive since such a Universe 
would have been very old at the time of Big Bang Nucleosynthesis (BBN) 
suggesting that all neutrons have decayed and are unavailable to be 
incorporated in \4he.  In fact, as Sethi \etal correctly pointed out, 
the expansion rate is so slow that the weak reactions remain in equilibrium 
sufficiently long to permit a ``simmering" synthesis of the required amount 
of \4he.  However, such an old Universe also leaves more time to burn away 
D and \he3 so that no astrophysically significant amounts can survive.  The 
observations of deuterium in high-redshift, low-metallicity QSO absorbers
~\cite{bt}, the observations of lithium in very old, very metal-poor 
halo stars (the ``Spite plateau")~\cite{lith}, and those of helium in 
low-metallicity extragalactic \hii regions~\cite{4he} require an 
internally consistent primordial origin. The Sethi \etal claim that 
deuterium could have a non-primordial origin is without basis as shown 
long ago by Epstein, Lattimer \& Schramm~\cite{els}.  Nevertheless, the paper of Sethi 
\etal \cite{sethi} prompted us to reinvestigate primordial nucleosynthesis 
in those power-law cosmologies which may produce ``interesting" amounts of 
\4he so as to study the predicted yields for D, \he3, and \7li .

\section{Nucleosynthesis In Power-Law Cosmologies}
\begin{figure}[Fig1]
\centering
\leavevmode\epsfxsize=12cm \epsfbox{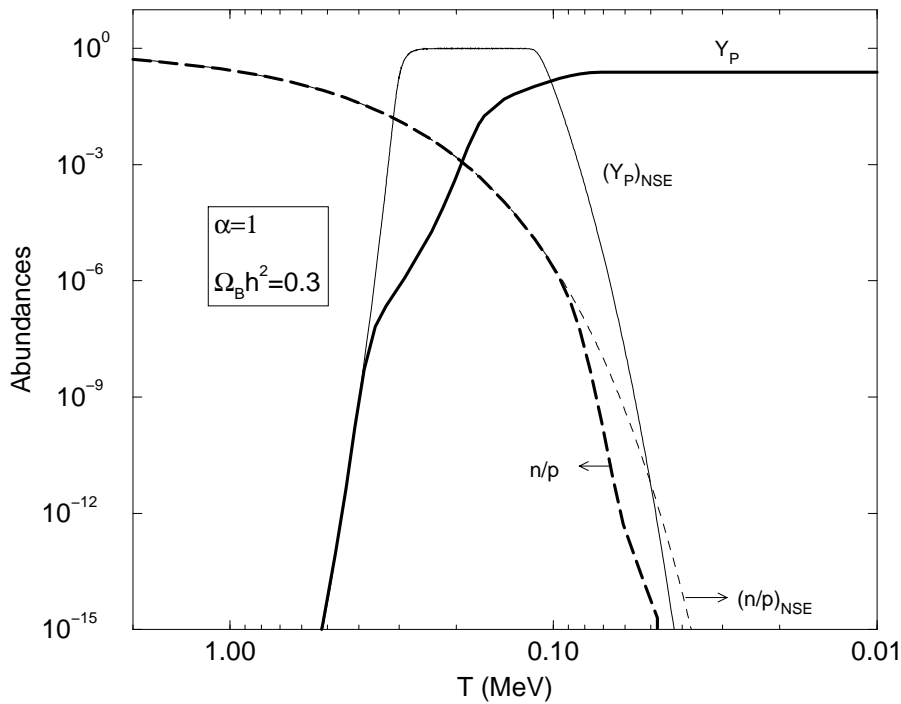}\\
\
\caption[Fig1]{\label{equib} Comparison of nucleosynthesis in the linear
expansion model (heavy curves) for the case of $\ypm = 0.24$ with the 
predictions of nuclear statistical equilibrium (lighter curves).  The 
solid curves are for the \4he mass fraction \yp, while the dashed curves 
show the evolution of the ratio of neutrons to protons.}
\end{figure}

{\it Preliminaries}.
For a power-law cosmology it is assumed that the scale factor varies
as a power of the age independent of the cosmological epoch:
\be
a/a_{0} = (t/t_{0})^{\alpha} = (1 + z)^{-1}, \label{abya0}
\ee
where the subscript `0' refers throughout to the present time and `$z$' 
is the redshift.  We may relate the present cosmic background radiation 
(CBR) temperature to that at any earlier epoch by $T = (1 + z)\beta T_{0}$, 
where $\beta\le 1$	accounts for any entropy production.  For the 
models we consider, $\beta =1$ after electron-positron annihilation.  
The Hubble parameter is then given by
\be
H=\frac{\dot{a}}{a}=\frac{\alpha}{t_0}\left(\frac{T}{\beta T_0}\right)^
{\frac{1}{\alpha}}. \label{hubble}
\ee
The second equality should be read with the understanding that it is 
not valid during the epoch of electron-positron annhilation due to the 
non-adiabatic nature of annhilations.  Power-law cosmologies with large 
$\alpha$ share the common feature that the slow Universal expansion rate 
permits neutrinos to remain in equilibrium until after electron-positron 
annihilation has ended so that neutrino and photon temperatures remain 
equal.  In this case the entropy factor for $T>m_e/3$ in eq.~\ref{hubble} 
is,
\be \beta= (29/43)^{1/3}
\ee
in contrast to the standard Big Bang nucleosynthesis (SBBN) value of 
$(4/11)^{1/3}$.  As \alp increases, the expansion rate at a fixed 
temperature decreases due to the dominant effect of the $1/\alpha$ power.  
Another useful way to view this is that at a fixed temperature, a power-law 
Universe with a larger $\alpha$ is older. As a consequence of the decreasing 
expansion rate, the reactions remain in equilibrium longer. In particular, 
as pointed out by Sethi \etal \cite{sethi} for the linear expansion model 
($\alpha=1$), the weak interactions remain in equilibrium to much lower 
temperatures than in the SBBN scenario, allowing neutrons and protons to 
maintain equilibrium at temperatures below 100 keV, as can be seen in Fig.
~\ref{equib}.  As is evident from Fig.~\ref{equib}, the \4he production rate 
below about $0.4~\mathrm{MeV}$ is too slow to maintain nuclear statistical 
equilibrium.  However, the presence of neutrons in equilibrium and the 
enormous amount of time available  for nucleosynthesis 
during neutron-proton equilibrium (compared to SBBN) make it possible to build up a significant 
abundance of \4he~\cite{sethi}.

\begin{figure}[Fig2]
\centering
\leavevmode\epsfxsize=12cm \epsfbox{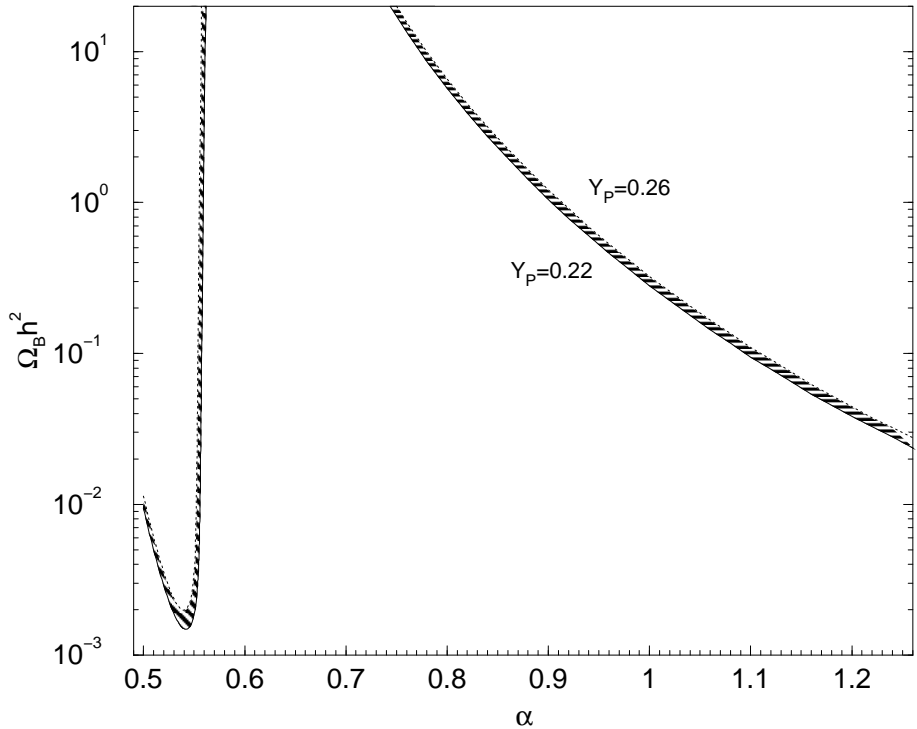}\\
\
\caption[Fig2]{\label{iso-he4} Iso-abundance contours of the \4he 
mass fraction (\yp) in the baryon density -- $\alpha$ plane.  
The shaded band corresponds to helium abundances in the range
$0.22 \leq \mathrm{Y}_\mathrm{P} \leq 0.26$.}
\end{figure}

The above discussion is not restricted to $\alpha=1$, but applies 
for all values of $\alpha$ which are sufficiently large (so that 
the expansion rate is sufficiently small) to allow neutrons to stay 
in equilibrium long enough to enable synthesis of \4he in sufficient 
amounts, as we show in Fig.~\ref{iso-he4}.  Although we explore a 
larger range in $\alpha$ in this paper, we present detailed results 
for $0.75 \le \alpha \le 1.25$, a range consistent with the age and 
expansion rate of the Universe, and we check these results for consistency 
with independent (\ie, non-BBN) constraints on the baryon density.  The 
iso-abundance contours in Fig.~2 show clearly that as $\alpha$ decreases 
towards 0.75, a larger baryon density is required to produce the same 
abundance of \4he.  For example, although $\ypm = 0.24$ can be synthesized 
in the $\alpha=0.75$ model, the density of baryons required is very large: 
$\omegab h^2\simeq 20$. These `large $\omegab$' models are constrained 
by dynamical estimates of the mass density, an issue we discuss later.

{\it Helium-4 abundance}.
In an earlier study~\cite{us}, we showed that there is a very small 
region, centered on $\alpha=0.55$, for which the light elements can 
be produced in abundances similar to those predicted by SBBN.  But, 
this small window is closed by the SNIa magnitude-redshift data~\cite{snia}.  
Here we are concerned with larger values of $\alpha$ and, correspondingly, 
larger baryon-to-photon ratios ($\eta$).  First we consider the 
nucleosynthesis of \4he in these models.  Figure~\ref{iso-he4} shows the 
connection between the baryon density ($\Omega_B h^{2} = \eta_{10}/273$, 
where $\eta_{10} \equiv 10^{10}n_{N}/n_{\gamma}$) and $\alpha$ set by the 
requirement that the primordial helium mass fraction lie in the generous 
range $0.22 \leq \ypm \leq 0.26$.  We have included in Fig.~\ref{iso-he4} 
the region investigated in Ref.~\cite{us}, $\alpha < 0.6$, as well.  To 
understand the features in Fig.~\ref{iso-he4}, we need to isolate the 
important factors controlling the synthesis of helium.  In SBBN, the \4he 
abundance is essentially controlled by the number density ratio of neutrons 
to protons ($n/p$) at the start of nucleosynthesis ($T=T_{BBN}\approx 
80$~keV). This ratio in turn is determined by (1) the $n-p$ ratio at 
`freeze-out' ($T=T_{f}$) of the neutron-proton interconversion rates which 
may be approximated by $\npf=\exp(-Q/T_f)$, where $Q=1.293~\mathrm{MeV}$ 
is the neutron-proton mass difference and, (2) the time available for 
neutrons to decay after freeze-out, $\dtdec=t(T_{BBN})-t(T_f)$.  In 
contrast, for power-law cosmologies another factor comes into play -- the 
time available for nucleosynthesis, $\Delta t_{BBN}$, before the nuclear 
reactions freeze-out.  For larger $\alpha$, the expansion rate of the 
Universe (at fixed temperature) is smaller and the Universe is older.  
Hence, for larger $\alpha$ neutrons remain in equilibrium longer and the 
freeze-out temperature ($T_f$) is smaller, so that $\npf$~is smaller.  
However, the effect of the increase in $\Delta t_{BBN}$ as $\alpha$ 
increases dominates that due to the change in $T_f$.  For $\alpha=0.50$ 
the freeze-out temperature is around 4~MeV whereas for $\alpha=0.55$, 
$T_f\simeq 1$~MeV which implies a decrease in $(n/p)_f$ by a factor 
of about 2.5.  On the other hand, the age of the Universe at $T=10$~keV 
(about the temperature when SBBN ends) is a factor of 25 larger for 
$\alpha = 0.55$ relative to that for $\alpha=0.50$.  Thus, for the same 
$\eta$, increasing $\alpha$ from 0.50 to 0.55 has the effect of increasing 
the $^4$He abundance because more time is available for nucleosynthesis.  
But, since decreasing the baryon density decreases the nuclear reaction 
rates leading to a decrease in $^4$He, we may understand the trend of 
the smaller baryon density requirement as $\alpha$ increases from 0.50 
to about 0.55, even though the decrease in $T_f$ opposes this effect.  
The time-delay between `freeze-out' and BBN, $\dtdec$, which has, 
until now, been much smaller than $\tau_n$, becomes comparable to 
it at $\alpha\sim 0.55$.  Since a larger $\alpha$ results in an older 
Universe at a fixed temperature, $\dtdec$~increases with $\alpha$.  
Thus for $\alpha \ga 0.55$, \yp~is increasingly suppressed (exponentially) 
as $\alpha$ is increased.  The only way to compensate for this is by 
increasing $T_{BBN}$ (since $\dtdec \propto (T_{BBN})^{-1/\alpha}$), 
which may be achieved by increasing the baryon density.  But since 
$T_{BBN}$ depends only logarithmically on the baryon density~\cite
{kolb+turner}, this accounts for the exponential rise in the required 
value of $\Omega_B h^2$ as $\alpha$ increases. This trend cannot 
continue indefinitely; the curve must turn over for reasons we 
describe below. 

From Fig.~\ref{iso-he4}, it is apparent that in the ``large $\alpha$" 
range, the required value of $\omegab h^2$ decreases with increasing 
$\alpha$.  In our previous analysis~\cite{us} of \4he nucleosynthesis 
which concentrated on $\alpha$ in the vicinity of 0.55, we implicitly 
assumed that the age of the Universe at $T = T_f$ was not large enough 
for appreciable amounts of \4he to have been built up. This assumption 
breaks down for large values of $\alpha$ {\it and} $\eta$.  Since D, 
\he3 and \h3 are not present in appreciable quantities, a large value 
of $\eta$ is needed to boost the \4he production rate.  Now, the larger 
the value of $\alpha$, the longer neutrons remain in equilibrium, thus 
allowing more \4he to be slowly built up, with the neutrons incorporated 
in \4he being replaced via $p\rightarrow n$ reactions.  Roughly speaking, 
the required value of $\eta$ for a given $\alpha$ is set by the condition:
\begin{equation}
\left[\deriv{\ypm}{t}\right]_{T=T_f} \sim\;\; 0.24/t(T_f)\,. \label{req} 
\end{equation}    
The effects of $\alpha$ on $t(T_f)$ and $\eta$ on $\mathrm{d}\ypm/
\mathrm{d}t$ complement each other, giving rise to the trend shown 
by the \4he iso-abundance curves in Fig.~\ref{iso-he4} for $\alpha 
\ga 0.75$.

\begin{figure}[Fig3]
\centering
\leavevmode\epsfxsize=12cm \epsfbox{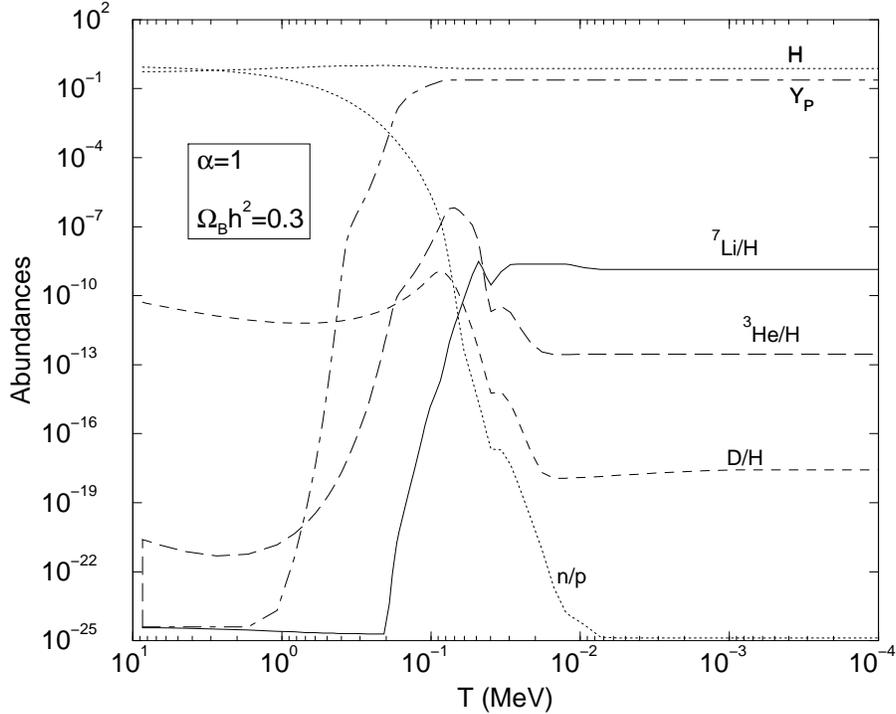}\\
\
\caption[Fig3]{\label{alpha1} Evolution of the light element 
abundances as a function of the photon temperature in an $\alpha 
= 1$ Universe. }
\end{figure}

\begin{figure}[Fig4]
\centering
\leavevmode\epsfxsize=12cm \epsfbox{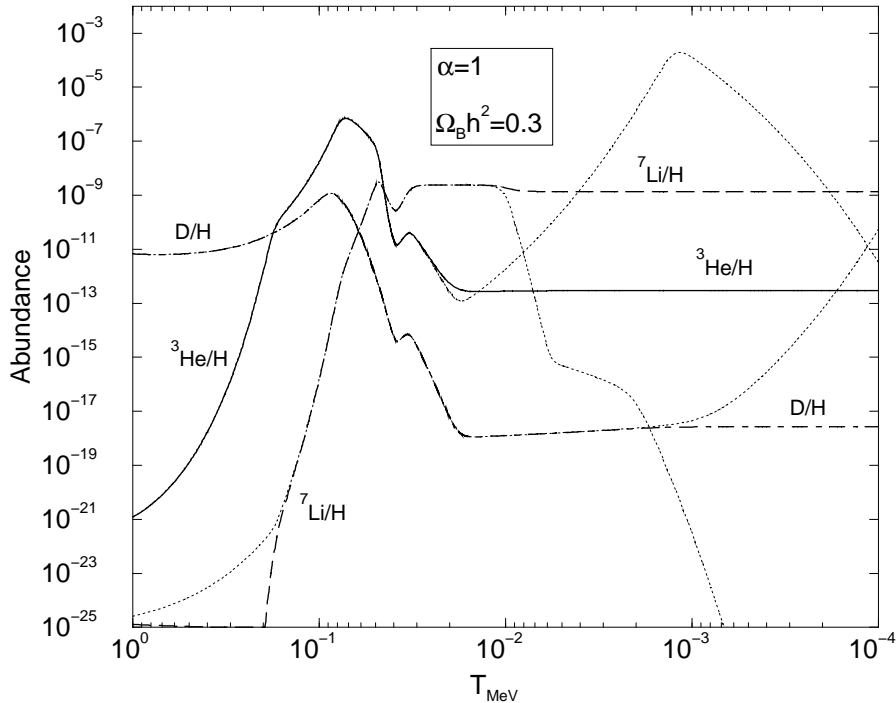}\\
\
\caption[Fig4]{\label{analytic} Comparison of the light element 
abundances with their steady state values as a function of the 
photon temperature in an $\alpha = 1$ Universe. The dotted curves 
correspond to the equilibrium solution.}
\end{figure}

{\it Light element abundances in the linear expansion model}.
We now turn to the production of deuterium and \he3. For large 
$\alpha$ (\eg, $\alpha=1$), we expect the deuterium abundance 
to be insignificant since D can be efficiently burned to \he3   
during the long time available for nucleosynthesis.  The mean 
lifetime of deuterium against destructive collisions with protons 
at a low temperature of 10 keV is around 3 days; at this temperature 
the $\alpha=1$ Universe is already 300 years old!  The fact that 
the timescales are so different allows us to derive analytical 
expressions for the deuterium, helium-3 and lithium-7 (beryllium-7)
mass fractions (to be denoted by $X_D$, $X_3$ and $X_7$ respectively). 
The generic equation for the rate of evolution of the mass fraction 
of nuclide ``$a$" can be parameterized as
\begin{equation}
\deriv{X_a}{t}=R_\mathrm{prod}(a)-R_\mathrm{dest}(a)X_a,
\end{equation}
where ``prod'' and ``dest'' refer to the production and destruction 
rates of nuclide ``$a$''. Given that the universe remains at the same 
temperature for a very long time (compared to the reaction time scales), 
it is not surprising that $X_a$ achieves its steady-state value 
at each temperature (for a detailed discussion in the context of SBBN, 
see \cite{esmailzadeh}),
\begin{equation}
X_a\approx \frac{R_\mathrm{prod}(a)}{R_\mathrm{dest}(a)}.
\label{xdx3x7}
\end{equation}
We can write this explicitly for the simplest case -- deuterium:
\begin{equation}
X_D = 2\;\frac{\left(\Gamma_{np} +\Gamma_{pp}/2\right)X_p}
{\Gamma_{pD} +\Gamma_{\gamma D}}
\label{xd}
\end{equation}
where the various $\Gamma$s represent the relevant deuterium creation
($n+p\longrightarrow D+\gamma$ and $p+p\longrightarrow D+e^++\nu$) rates 
per target proton, and destruction ($D(p,\gamma)$\he3 and $D(\gamma,p)n$) 
rates per target deuterium.  All of these rates can be obtained from 
Ref.~\cite{FCZII}.  Once the reaction rates become smaller than the universal 
expansion rate (say at some temperature $T_\star$), the abundances freeze 
out with values close to $X_a$ at the corresponding $T_\star$.  This is 
illustrated in Fig.~\ref{analytic} which clearly shows that the steady-state 
solution works very well. We note here that the steady state (dotted) curves 
in Fig.~\ref{analytic} are not independent analytic derivations, but use 
the abundances of the various nuclei as calculated by the numerical code.  
The figure intends to emphasize that nucleosynthesis in this (linear 
expansion) model can be well represented by the steady-state solutions 
in eq.~\ref{xdx3x7}. 
In the expression for $X_D$ (see eq.~\ref{xd}), the $n+p$ reaction term 
dominates until about 20 keV after which the $p+p$ reaction makes the 
dominant contribution. The final deuterium abundance is thus determined 
by the weak pp reaction ($p+p\longrightarrow D+e^++\nu$), the effect of 
which can be seen in Fig.~\ref{alpha1} as the very slow rise in $X_D$ 
between temperatures of 10 keV and 1 keV (at which point the D abundance 
freezes out).  Since both $^3$He and $^7$Li freeze out much earlier, 
they do not get any significant boost from the weak pp reaction. 

From eq.~\ref{xd}, $X_D$, and thus $X_3$ ($^3$He is formed from $D$), are 
proportional to $X_n$, the neutron abundance.  One striking feature in 
Fig.~\ref{alpha1} is the boost to the neutron  abundance (and hence the 
abundances of D and $^3$He) at temperatures around 40 keV.  The effect 
is subtle and may be missed in BBN codes with a limited nuclear reaction 
network. The slow rate of expansion of the universe during nucleosynthesis 
facilitates the production of a relatively large ``metal" (A$\geq8$) 
abundance ($X_\mathrm{metals}\simeq 3\times10^{-7}$).  In particular, 
$^{13}$C is produced in these models through the chain: $^{12}$C+p$
\longrightarrow$ $^{13}$N+$\gamma$ and the subsequent beta-decay of 
$^{13}$N.  In this environment 
$^{13}$C+$^4$He$\longrightarrow$$^{16}$O+$n$ leads to the production 
of free neutrons.  
   
The mass-7 abundance is entirely due to the production of $^7$Be 
through the reaction $^3$He+$^4$He$\longrightarrow$$^7$Be+$\gamma$.
$^7$Be decays to $^7$Li by electron capture once the universe has 
cooled sufficiently to permit the formation of atoms. Once formed, 
it is difficult to destroy $^7$Be at temperatures $\la 100$ KeV. 
In contrast, $^7$Li is very easily destroyed, specifically through 
its interaction with protons.  Since the $^7$Be production (and 
thus \7li) follows the evolution of the $^3$He abundance, and there 
is very little destruction of $^7$Be, $^7$Li also benefits from the 
boost to the neutron abundance described in the last paragraph.
This has the effect of boosting the \7li abundance from $10^{-11}$ 
(if this source of neutrons were not included) to $10^{-9}$.  This
is significant in that, at the level of a few parts in 10$^{10}$ 
(\eg,~\cite{osw}), the primordial lithium abundance lies between 
these two estimates.

\begin{figure}[Fig5]
\centering
\leavevmode\epsfxsize=12cm \epsfbox{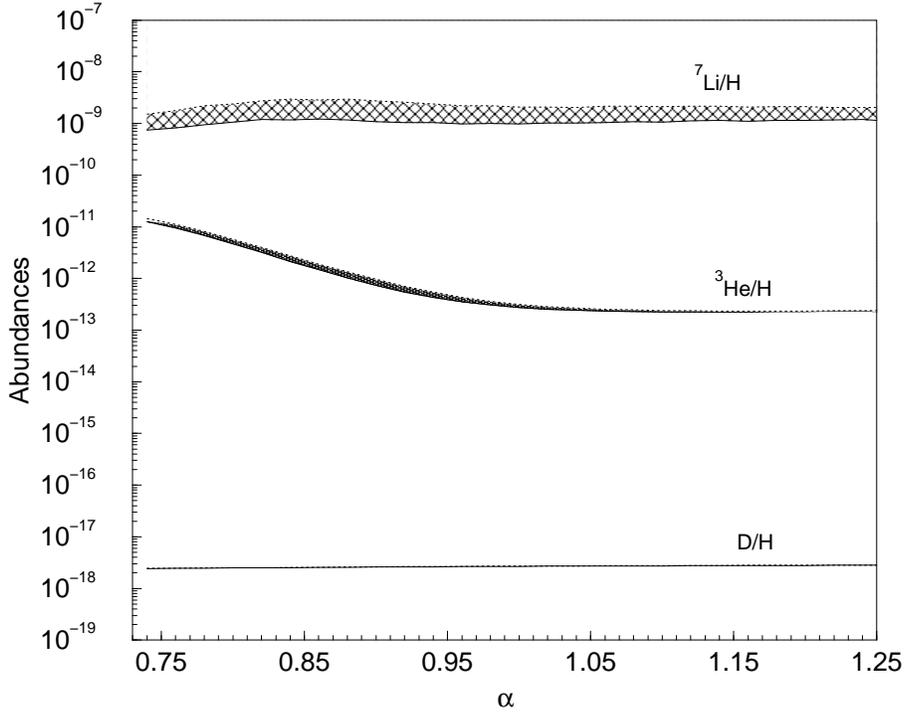}\\
\
\caption[Fig5]{\label{abund} Abundances of \he3, 
D and \7li in power-law cosmologies for different values of the 
expansion index ($\alpha$).  The shaded bands correspond to helium
abundances in the range $0.22 \leq \mathrm{Y}_\mathrm{P} \leq 0.26$.
}
\end{figure}

{\it Light element abundances vs. $\alpha$}. 
Having explored BBN in the linear model ($\alpha = 1$) it is now 
important to ask how these results depend on $\alpha$.  It is clear 
from Fig.~\ref{abund} that nothing dramatically different happens 
as $\alpha$ changes; this is simply because the key physics remains 
the same.  In preparing Fig.~\ref{abund} we adjust the value of $\eta$ 
(baryon density) for each choice of $\alpha$ so that the primordial 
\4he mass fraction lies between 22\% and 26\%.  As $\alpha$ increases, 
the nuclei freeze out at lower temperatures since the expansion rate 
at the same temperature is lower for a larger $\alpha$.  The effect 
of this can be gauged by the behavior of $X_D$, $X_3$ and $X_7$ with 
respect to temperature, as given by eq.~\ref{xdx3x7}.  For deuterium 
this implies a small increase with $\alpha$ due to the pp (weak) 
reaction, which is also reflected in the behavior of $^3$He for 
$\alpha\ga 1$. The fall of $^3$He with increasing $\alpha$ for 
$\alpha\la 1$ is due to larger destruction of $^3$He because of 
the increase in the time available for the nuclear reactions.  As 
already mentioned, the abundance of $^7$Be depends critically on 
the evolution of the $^3$He abundance; so while the mass-7 ($^7$Be) 
abundance increases appreciably with the increase in baryon density, 
it is relatively unaffected by a change in $\alpha$. 

Note that in those power-law models which can simultaneously reproduce
an acceptable \4he abundance along with a consistent age and expansion 
rate, the corresponding baryon density must be very large, $0.04\le 
\omegab h^2\le 6.4$ ($11\le \eta_{10}\le 1750$; see Fig.~2). Most -- if 
not all -- of this range is far too large for consistency with independent 
(non-BBN) estimates of the universal density of baryons ($\eta \la 7.4$~
\cite{fhp}) or, for that matter, the total matter density~\cite{neta}.  
Conservatively, clusters limit the total (gravitating) matter density 
to $\Omega_M \la 0.4$ so that if there were no non-baryonic dark matter, 
$\omegab h^2\la 0.2$ ($\eta\la 54$) for $h\sim 0.7$.  However, if the 
X-ray emission from clusters is used to estimate the cluster baryon 
fraction (see~\cite{shf}), the universal baryon density should be 
smaller than this very conservative estimate by a factor of 7-8 
(consistent with the upper bound from the baryon inventory of Fukugita, 
Hogan and Peebles~\cite{fhp}).  Thus power-law cosmologies constrained 
to reproduce \4he (only), an acceptable age and magnitude-redshift 
relation, and an acceptable baryon density, must have $\alpha$ restricted 
to a very narrow range: $1\la \alpha \la  1.2$.  Furthermore, the baryon 
density in even this restricted range is large when compared with 
estimates~\cite{shf} of the baryon density from cluster X-rays.  
Finally, for $\alpha$ in the narrow range of $1\la \alpha \la  1.2$ 
and $0.22\le\ypm\le 0.26$, the other light element abundances are 
restricted to \7li/H $> 10^{-9}$, \he3/H $< 3\times 10^{-13}$, and D/H 
$< 3\times 10^{-18}$.  For deuterium and helium-3 this is in very strong
disagreement (by 8 to 13 orders of magnitude!) with observational data 
(for a review see~\cite{osw}).  Although the predicted $^7$Li abundance
is comparable to that observed in the solar system, the local ISM, and 
in \popi stars, it is larger than the primordial abundance inferred 
from the \popii halo stars~\cite{lith,osw}, and marginally 
inconsistent with the observations of lithium in the ISM of the LMC 
~\cite{gs}.

\section{Conclusions}
In response to the claim~\cite{sethi} that a power-law Universe
expanding linearly with time could be consistent with the constraints
on BBN, we have reexamined these models.  Although it is true that
observationally consistent amounts of \4he can be produced in these
models, this is not the case for the other light elements D, \he3, 
\7li.  Furthermore, consistency with \4he at $\alpha = 1$ requires 
a very high baryon density ($75\le \eta_{10} \le 86$ or $0.27\le 
\omegab h^2 \le 0.32$), inconsistent with non-BBN estimates of the 
universal baryon density and, even with the total mass density.  We 
have also investigated BBN in power-law cosmologies with $\alpha > 
1$ and have confirmed that although the correct \4he abundance can 
be produced, the yields of the other light elements D, \he3, and \7li 
are inconsistent with their inferred primordial abundances.  In general, 
power-law cosmologies are unable to account simultaneously for the 
early evolution of the Universe (BBN) (which requires $\alpha \simeq 
0.55$) and for its presently observed expansion (which requires $\alpha 
= 1\pm0.2$).

\acknowledgments
 
This work was supported at Ohio State by DOE grant DE-AC02-76ER01545.

\end{document}